\documentclass[10pt, conference, letterpaper]{IEEEtran}
\usepackage[letterpaper, left=0.65in, right=0.65in, bottom=1.01in, top=0.75in]{geometry}
\IEEEoverridecommandlockouts
\usepackage{cite}
\usepackage{amsmath,amssymb,amsfonts}
\usepackage{graphicx}
\usepackage[caption=false,font=footnotesize]{subfig}
\usepackage{textcomp}
\usepackage{xcolor}
\usepackage{tabularx}
\usepackage{multirow}
\usepackage{diagbox}
\usepackage[binary-units]{siunitx} 
\DeclareSIUnit{\bps}{bps}
\usepackage{flushend}
\usepackage{amsthm}

\begin{document}
\title{A Review of Multiple Access Techniques for Intelligent Reflecting Surface-Assisted Systems}

\author{\IEEEauthorblockN{Wei Jiang\IEEEauthorrefmark{1} and Hans D. Schotten\IEEEauthorrefmark{1}\IEEEauthorrefmark{2}}
\IEEEauthorblockA{\IEEEauthorrefmark{1}Intelligent Networking Research Group, German Research Center for Artificial Intelligence (DFKI), Germany\\
  }
\IEEEauthorblockA{\IEEEauthorrefmark{2}Department of Electrical and Computer Engineering, University of Kaiserslautern (RPTU), Germany\\
 }
}
\maketitle

\begin{abstract}
Intelligent Reflecting Surface (IRS) is envisioned to be a technical enabler for the sixth-generation (6G) wireless system. Its potential lies in delivering high performance while maintaining both power efficiency and cost-effectiveness. Previous studies have primarily focused on point-to-point IRS communications involving a single user. Nevertheless, a practical system must serve multiple users simultaneously. The unique characteristics of IRS, such as non-frequency-selective reflection and the necessity for joint active/passive beamforming, create obstacles to the use of conventional multiple access (MA) techniques.  This motivates us to review various MA techniques to make clear their functionalities in the presence of IRS. Through this paper, our aim is to provide researchers with a comprehensive understanding of challenges and available solutions, offering insights to foster their design of efficient multiple access for IRS-aided systems.
\end{abstract}

\section{Introduction}

Intelligent reflecting surface (IRS) has recently captured a lot of attention \cite{Ref_renzo2020smart}. Essentially, IRS is a surface composed of numerous small and cheap reflecting elements. Each of these elements can adaptively rotate the phase of an incoming signal. In contrast to traditional wireless techniques in conventional cellular systems \cite{Ref_jiang2024TextBook} that passively adjust to a radio channel, IRS offers a smart propagation environment by actively modifying the channel \cite{Ref_jiang20236GCH7}. Through controllable reflection, the reflected signals can be combined constructively to enhance signal strength or destructively to mitigate interference. As the reflecting elements, such as printed dipoles or positive-intrinsic-negative (PIN) diodes, merely passively reflect signals, there is no need for radio-frequency chains for signal transmission and reception. Even for the recently proposed paradigm called active IRS \cite{Ref_zhang2023activeRIS}, the use of low-power amplifiers makes this technique still power- and cost-effective, compared to active antenna arrays. This characteristic positions the IRS as a technological candidate for the sixth-generation (6G) system \cite{Ref_jiang2021road}. 

A large and expanding body of literature has delved into various aspects related to building IRS-aided wireless systems, e.g., the reflection optimization \cite{Ref_wu2019intelligent}, estimation of cascaded channels \cite{Ref_wang2020channel}, practical constraints \cite{Ref_jiang2023performance}, and the use of active elements \cite{Ref_zhang2023activeRIS}. Prior studies primarily focused on point-to-point IRS communications, which involve a single user, a base station, and a surface. However,  a practical wireless system needs to serve many users simultaneously, raising the problem of multiple access (MA). Like the leap from single-user MIMO to multi-user MIMO \cite{Ref_jiang20236GCH8}, a lot of challenges have to be addressed for multi-user IRS communications.  The incorporation of IRS, owing to its transformative capability in reconfiguring wireless channels and considering particular hardware limitations, poses barriers to multi-user signal transmission. For example, the lack of frequency-selective reflection, where the phase shift of each element cannot vary across frequency, results in incompatibility for widely adopted frequency-division systems. By far, which MA techniques are efficient in the presence of IRS is still not clear. This motivates us to conduct a comparative study, offering a holistic view of the current state of the art.

In this paper, we aim to review different MA techniques to make clear their behaviors in the presence of IRS. Their fundamentals are comparatively presented through mathematical representations of transmitted and received signals, alongside the derivation of closed-form expressions for achievable spectral efficiency and sum rate. Due to differences in system modeling, such as single-antenna vs. multi-antenna base stations, single vs. multiple IRS surfaces, and the presence or absence of direct links, it is challenging to create a unified presence. Therefore, we distill the key ideas of different MA approaches and adapt them to a unified system model.  The primary contribution of this paper lies in offering an exhaustive survey, presenting a unified illustration, and providing a holistic view. Through this paper, we aim to provide researchers with a comprehensive understanding of challenges and available solutions, offering insights to foster their design of efficient multiple access for IRS-aided systems. 


\section{State of the Art}

Through an exhaustive literature review, we can offer a summary of the existing methods as follows: the simplest way is to assign users to orthogonal time-frequency resource pieces by time-division multiple access (TDMA) \cite{Ref_zheng2020intelligent_COML} or (orthogonal) frequency-division multiple access (FDMA/OFDMA) \cite{Ref_jiang2023userscheduling}. TDMA suggests a discontinuous transmission for a user, streamlining system operation by allowing certain tasks like passive beamforming optimization to be carried out in the time slots allocated to other users. FDMA remains relevant in specific 6G scenarios, notably in IRS-assisted Internet of Things applications \cite{Ref_chu2022resource} due to its simple implementation. Simultaneously, intelligent surfaces can seamlessly integrate into existing OFDMA-based networks to enhance performance, particularly LTE and 5G NR environments. However, TDMA and FDMA are orthogonal multiple access (OMA) schemes, which are inefficient since each user only utilizes a small portion of available resources. Therefore, the researchers investigated the use of non-orthogonal multiple access (NOMA) in IRS-aided systems, where its achievable performance was compared with that of OMA \cite{ Ref_zheng2020intelligent_COML, Ref_jiang2023orthogonal}. The authors of \cite{Ref_ding2020simple} provided a simple IRS-NOMA design. In addition, space-division multiple access (SDMA) was considered for large antenna arrays or high frequencies \cite{Ref_jiang2024beam, Ref_jiang2022dualbeam}. Globally optimized reflection is achieved by joint passive/active optimization using semidefinite relaxation (SDR) \cite{Ref_wu2019intelligent, Ref_jiang2023capacity}, whereas it is only tractable in a single-user setup. Selecting a single user opportunistically to transmit at each time substantially simplifies the system design. Based on the availability of channel state information (CSI), two variants, i.e., user selection  \cite{Ref_gan2021user, Ref_jiang2023userSelection} and opportunistic beamforming \cite{Ref_nadeem2021opportunistic, Ref_dimitropoulou2022opportunistic}, have been designed. Nevertheless, these two schemes suffer from performance degradation due to inefficient single-user transmission. In \cite{Ref_jiang2023opportunistic, Ref_jiang2023simple}, new methods that allow all users for simultaneous transmission while only selecting an opportunistic user for optimal reflection optimization have been proposed. In addition, the work in \cite{Ref_jiang2022multiuser} studied the effect of discrete phase shifts on multi-user IRS communications.  We have also analyzed and compared different MA techniques for IRS-aided vehicular networks \cite{Ref_jiang2022intelligent}.

\section{Unified System Model}
\begin{figure}[!t]
    \centering
    \includegraphics[width=0.42\textwidth]{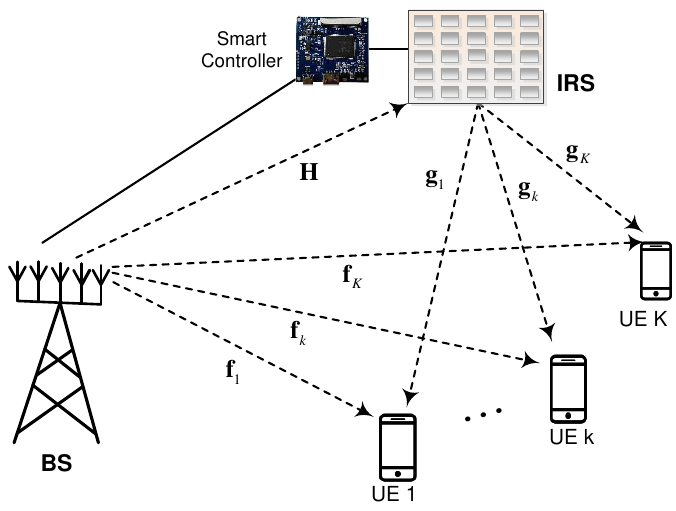}
    \caption{Illustration of a multi-user IRS system, consisting of a base station with $M$ antennas, $K$ users with single-antenna UE, and an IRS surface incorporating $N$ elements.  }
    \label{diagram:system}
\end{figure}

As illustrated in \figurename \ref{diagram:system}, this paper considers an IRS-aided system, wherein a surface equipped with $N$ reflecting elements is deployed to enhance communications between an $M$-antenna base station (BS) and $K$ single-antenna user equipment (UE). Due to the page limit, we only focus on the downlink transmission while extending the discussed approaches and analyses to the uplink is straightforward. We use an $M\times 1$ vector
\begin{align} \label{EQN_f_k}
    \mathbf{f}_{k}=\left[f_{k1},f_{k2},\ldots,f_{kM}\right]^T
\end{align}
to model the channel between the $k^{th}$ UE and the BS, where $f_{km}$, $\forall m=1,\ldots,M$ and $\forall k=1,\ldots,K$, stands for the channel gain between the $m^{th}$ BS antenna and the $k^{th}$ UE. The channel between the IRS and the $k^{th}$ UE is defined as
\begin{align} 
    \mathbf{g}_{k}=\left[g_{k1},g_{k2},\ldots,g_{kN}\right]^T,
\end{align}
where $g_{kn}$, $\forall n=1,\ldots,N$ and $\forall k=1,\ldots,K$, represents the channel gain between the $n^{th}$ reflecting element and the $k^{th}$ UE. Let $h_{nm}$, $\forall n=1,\ldots,N$ and $\forall m=1,\ldots,M$, express the channel gain between the $n^{th}$ reflecting element and the $m^{th}$ antenna. Then, the channel from the BS to the $n^{th}$ element is $\mathbf{h}_{n}=[h_{n1},h_{n2},\ldots,h_{nM}]^T$, the channel matrix from the BS to the IRS is expressed as $\mathbf{H}\in \mathbb{C}^{N\times M}$, where the $n^{th}$ row of $\mathbf{H}$ equals to $\mathbf{h}_n^T$. 

A smart controller on the surface establishes a wired or wireless connection with the BS. Its function involves dynamically adjusting the phase shift of each reflecting element based on acquired channel state information (CSI) through periodic channel estimation. For ease of illustration, we assume that the BS perfectly knows the CSI, as did most works, while the estimation methods can refer to prior works like \cite{Ref_wang2020channel}. The reflection of a typical element indexed as $n$ is mathematically represented by  $\epsilon_{n}=a_{n} e^{j\phi_{n}}$. Here, $\phi_{n}$ represents the phase rotation within the range of $[0,2\pi)$, and $a_{n}\in [0,1]$ denotes the amplitude attenuation. According to \cite{Ref_wu2019intelligent}, $a_{n}=1$ is identified as the optimal attenuation, maximizing the received signal strength and simplifying implementation complexity. Consequently, the focus of reflection optimization lies solely on $\phi_{n}$. This analysis excludes hardware impairments, such as discrete phase shifts \cite{Ref_wu2020beamforming} and phase noise \cite{Ref_jiang2023performance}. 

When the BS sends a vector of transmitted symbols as $\mathbf{x}=[x_1,\ldots,x_M]^T$ under the power constraint of $P_d$, the $k^{th}$ user observes the received signal of
\begin{equation} \label{eqMystemModel}
    y_k=\sqrt{P_d}\Biggl( \sum_{n=1}^N g_{kn} e^{j\phi_{n}} \mathbf{h}_{n}^T + \mathbf{f}_{k}^T \Biggr) \mathbf{x} + n_k,
\end{equation}
where $n_k$ denotes additive white Gaussian noise (AWGN) with zero mean and variance $\sigma_n^2$, i.e., $n_k\sim \mathcal{CN}(0,\sigma_n^2)$. Define a diagonal phase-shift matrix as $\boldsymbol{\Theta}=\mathrm{diag}\{[e^{j\phi_{1}},\ldots,e^{j\phi_{N}}]\}$, \eqref{eqMystemModel} can be rewritten in matrix form as
\begin{equation} \label{EQN_IRS_RxSignal_Matrix}
    y_k= \sqrt{P_d}\Bigl(\mathbf{g}_k^T \boldsymbol{\Theta} \mathbf{H} +\mathbf{f}_k^T\Bigr)\mathbf{x} +n_k.
\end{equation}

\section{Comparative Illustration of IRS Multiple Access Techniques}

This section presents a comprehensive understanding of various MA techniques in IRS-aided systems, encompassing TDMA, FDMA/OFDMA, SDMA, NOMA, user selection, and opportunistic beamforming. Their distinctive characteristics are comparatively explained, and closed-form expressions for per-user spectral efficiency (SE) and sum rate are derived.

\subsection{TDMA}  In this approach, the signaling dimensions along the time axis are divided into orthogonal time slots \cite{Ref_zheng2020intelligent_COML}. Each user utilizes the entire bandwidth but cyclically accesses the specific slot assigned to them. For simplicity, let's consider a radio frame that is divided into $K$ slots. According to \cite{Ref_zhang2018space}, the switching frequency of reflecting elements, fabricated by PIN diodes, can reach \SI{5}{\mega\hertz}. This equates to a switching time of \SI{0.2}{\micro\second}, significantly shorter than the usual channel coherence time, which typically falls in the range of milliseconds. It implies that the set of phase shifts can be adjusted specifically for the active user during each time slot. We denote the \textit{time-selective} phase shifts for user $k$ as $\boldsymbol{\Theta}_k=\mathrm{diag}\{[e^{j\phi_{1}[k]},\ldots,e^{j\phi_{N}[k]}]\}$, where $\boldsymbol{\Theta}_k\neq \boldsymbol{\Theta}_{k'}$, when $k\neq k'$.

The BS applies linear beamforming $\mathbf{w}_k\in \mathbb{C}^{M\times 1}$, where $\|\mathbf{w}_k\|^2\leqslant 1$, to transmit the signal intended for user $k$ at the $k^{th}$ slot. The information-bearing symbol $s_k$ is zero mean and unit-variance, i.e., $\mathbb{E}\left[|s_k|^2\right]=1$.  Substituting $\mathbf{x}=\mathbf{w}_k s_k$ into \eqref{EQN_IRS_RxSignal_Matrix}, we obtain
\begin{equation} 
    y_k= \sqrt{P_d}\Bigl(\mathbf{g}_k^T \boldsymbol{\Theta}_k \mathbf{H} +\mathbf{f}_k^T\Bigr)\mathbf{w}_k s_k +n_k.
\end{equation}
Through the joint optimization of active and passive beamforming, the instantaneous signal-to-noise ratio (SNR) can be maximized, leading to the optimization formula 
\begin{equation}  
\begin{aligned} \label{eqnIRS:optimizationMRTvector}
\max_{\boldsymbol{\Theta}_k,\:\mathbf{w}_k}\quad &  \biggl|\Bigl(\mathbf{g}_k^T \boldsymbol{\Theta}_k \mathbf{H} +\mathbf{f}_k^T\Bigr)\mathbf{w}_k\biggr|^2\\
\textrm{s.t.} \quad & \|\mathbf{w}_k\|^2\leqslant 1\\ 
  \quad & \phi_{n}[k]\in [0,2\pi), \: \forall n=1,\ldots,N, \forall k=1,\ldots,K,
\end{aligned}
\end{equation}
which is non-convex because the objective function is not jointly concave with respect to $\boldsymbol{\Theta}_k$ and $\mathbf{w}_k$.  Alternating optimization can be used to solve this problem by alternately optimizing $\boldsymbol{\Theta}_k$ and $\mathbf{w}_k$ \cite{Ref_wu2019intelligent}. The maximal-ratio beamforming $\mathbf{w}_{k}^{(0)}=\mathbf{f}_{k}^*/\|\mathbf{f}_{k}\|$ for the direct link can be applied as the initial value \cite{Ref_jiang2022intelligent}.
Thus, \eqref{eqnIRS:optimizationMRTvector} is simplified to 
\begin{equation}  \label{eqnIRS:optimAO}
\begin{aligned} \max_{\boldsymbol{\Theta}_k}\quad &  \biggl|\Bigl(\mathbf{g}_k^T \boldsymbol{\Theta}_k \mathbf{H} +\mathbf{f}_{k}^T\Bigr)\mathbf{w}_k^{(0)}\biggr|^2\\
\textrm{s.t.}  \quad & \phi_{n}[k]\in [0,2\pi), \: \forall n=1,\ldots,N, \forall k=1,\ldots,K.
\end{aligned}
\end{equation}
The objective function is still non-convex but it enables a closed-form solution by applying the triangle inequality
\begin{equation} 
    \biggl|\Bigl(\mathbf{g}_k^T \boldsymbol{\Theta}_k \mathbf{H} +\mathbf{f}_{k}^T\Bigr)\mathbf{w}_k^{(0)}\biggr| \leqslant \biggl|\mathbf{g}_k^T \boldsymbol{\Theta}_k \mathbf{H} \mathbf{w}_k^{(0)}\biggr| +\biggl|\mathbf{f}_{k}^T\mathbf{w}_k^{(0)}\biggr|. 
\end{equation}
The equality is achieved if and only if 
\begin{equation}
    \arg\left (\mathbf{g}_k^T \boldsymbol{\Theta}_k \mathbf{H} \mathbf{w}_k^{(0)}\right)= \arg\left(\mathbf{f}_{k}^T\mathbf{w}_k^{(0)}\right)\triangleq \varphi_{0k},
\end{equation}
where $\arg(\cdot)$ denotes the angle of a complex vector or scalar.

Define $\mathbf{q}_k=\left[q_{1k},q_{2k},\ldots,q_{Nk}\right]^H$ with $q_{nk}=e^{j\phi_{n}[k]}$ and  $\boldsymbol{\chi}_k=\mathrm{diag}(\mathbf{g}_k^T)\mathbf{H}\mathbf{w}_k^{(0)}\in \mathbb{C}^{N\times 1}$, we have $\mathbf{g}_k^T \boldsymbol{\Theta}_k \mathbf{H} \mathbf{w}_k^{(0)}=\mathbf{q}_k^H\boldsymbol{\chi}_k\in \mathbb{C} $.
Ignore the constant term $\bigl|\mathbf{f}_{k}^T\mathbf{w}_k^{(0)}\bigr|$, \eqref{eqnIRS:optimAO} is transformed to
\begin{equation}  \label{eqnIRS:optimizationQ} 
\begin{aligned} \max_{\boldsymbol{\mathbf{q}_k}}\quad &  \Bigl|\mathbf{q}_k^H\boldsymbol{\chi}_k\Bigl|\\
\textrm{s.t.}  \quad & |q_{nk}|=1, \: \forall n=1,\ldots,N, \forall k=1,\ldots,K,\\
  \quad & \arg(\mathbf{q}_k^H\boldsymbol{\chi}_k)=\varphi_{0k}.
\end{aligned}
\end{equation}
The solution for \eqref{eqnIRS:optimizationQ} can be derived as 
\begin{equation} \label{eqnIRScomplexityQ}
    \mathbf{q}^{(1)}_k=e^{j\left(\varphi_{0k}-\arg(\boldsymbol{\chi}_k)\right)}=e^{j\left(\varphi_{0k}-\arg\left( \mathrm{diag}(\mathbf{g}_k^T)\mathbf{H}\mathbf{w}_k^{(0)}\right)\right)}.
\end{equation}
Given $    \boldsymbol{\Theta}^{(1)}_k=\mathrm{diag}\{\mathbf{q}^{(1)}_k\}$, the process is alternated to optimize $\mathbf{w}_k$. The BS can apply a maximal-ratio method to maximize the strength of a desired signal, resulting in the active beamformer
$\mathbf{w}_k^{(1)} = \frac{(\mathbf{g}_k^T \boldsymbol{\Theta}^{(1)}_k \mathbf{H} +\mathbf{f}_{k}^T)^H}{\|\mathbf{g}_k^T \boldsymbol{\Theta}_k^{(1)} \mathbf{H} +\mathbf{f}_{k}^T\|}$.
Following the first iteration,  $\boldsymbol{\Theta}^{(1)}_k$ and $\mathbf{w}_k^{(1)}$ serve as the initial input for the subsequent iteration. This iterative process continues until convergence is achieved, yielding the optimal active and passeive beamformer denoted by $\mathbf{w}_k^{\star}$ and $\boldsymbol{\Theta}_k^{\star}$. Consequently, the sum rate of the TDMA-based IRS system can be figured out as 
\begin{equation} \label{IRS_EQN_TDMA_SE}
    C_{tdma}=\sum_{k=1}^K\frac{1}{K}\log\left(1+\frac{P_d \left|\left(\mathbf{g}_k^T \boldsymbol{\Theta}_k^\star \mathbf{H} +\mathbf{f}_k^T\right)\mathbf{w}_k^\star \right|^2 }{\sigma_n^2} \right).
\end{equation}

\subsection{FDMA/OFDMA} 

Unlike TDMA, where the phase shifts of IRS can be adjusted in different slots, it encounters the challenge of \textit{non-frequency-selective} reflection. Each element induces a common phase shift across subchannels. Hence, the surface can be optimized maximally for a specific user at a time. Other users across different subchannels must adopt this common configuration, resulting in phase-unaligned reflection. This raises an issue of \textit{user scheduling}: determining which user should be selected for passive beamforming optimization.
Our work presented in \cite{Ref_jiang2023userscheduling} has uncovered an interesting finding: when there are a limited number of users, user scheduling does not effectively improve performance. Furthermore, as the number of users increases, the utility of both user scheduling and reflection optimization becomes very marginal. In such cases, the phase shifts of IRS elements can be randomly set, denoted by a phase-shift matrix of $\boldsymbol{\Theta}_{r}$, each entry of which takes value randomly from $[0,2\pi)$. Given $\boldsymbol{\Theta}_{r}$, each user optimizes its active beamforming as $\mathbf{w}_{k}^{\divideontimes} = \frac{\left(\mathbf{g}_{k}^T \boldsymbol{\Theta}_{r} \mathbf{H} +\mathbf{f}_{k}^T\right)^H}{\left\|\mathbf{g}_{k}^T \boldsymbol{\Theta}_{r} \mathbf{H} +\mathbf{f}_{k}^T\right\|}$. In this case, the sum rate is 
 \begin{align}
     C_{fdma}&= \sum_{k=1}^K\frac{1}{K}\log\left(1+\frac{P_d \left|\left(\mathbf{g}_k^T \boldsymbol{\Theta}_{r} \mathbf{H} +\mathbf{f}_k^T\right)\mathbf{w}_k^{\divideontimes} \right|^2 }{\sigma_n^2} \right).
  \end{align}
Randomizing the phase shifts without CSI eliminates the requirement for channel estimation of cascaded channels, leading to a notable reduction in system complexity.


\subsection{NOMA} 

Despite its simplicity, OMA is inefficient due to under-utilization of resources. NOMA enables the service of multiple users over a resource unit. Multiple information symbols are superimposed into a composite waveform as $\mathbf{x}=\sum_{k=1}^K \sqrt{\alpha_k } \mathbf{w}_{k} s_k$, where  $\alpha_k$ represents the power coefficient subjecting to $\sum_{k=1}^K\alpha_k\leqslant 1$.
Typically, more power is assigned to users with lower channel gains to ensure detection reliability. A user with a stronger channel gain gets less power, but a reasonable SNR for correctly detecting its signal remains \cite{Ref_jiang2023orthogonal}. The surface can only be tuned to optimally aid a specific user $\hat{k}$ at a time with $\boldsymbol{\Theta}^{\star}_{\hat{k}}$ and $\mathbf{w}_{\hat{k}}^{\star}$, which can be derived using the alternating optimization as in TDMA. Other users have to share $\boldsymbol{\Theta}^{\star}_{\hat{k}}$ that is not favorable for them.  Given the combined channel gain $\mathbf{g}_{k}^T \boldsymbol{\Theta}^{\star}_{\hat{k}} \mathbf{H} +\mathbf{f}_{k}^T$, a user $k \neq \hat{k}$ can optimize its active beamforming as $\mathbf{w}_{k}^{\star} = \frac{(\mathbf{g}_{k}^T \boldsymbol{\Theta}^{\star}_{\hat{k}} \mathbf{H} +\mathbf{f}_{k}^T)^H}{\|\mathbf{g}_{k}^T \boldsymbol{\Theta}_{\hat{k}}^{\star} \mathbf{H} +\mathbf{f}_{k}^T\|}$.

At the receiver, successive interference cancellation (SIC) iteratively decodes the signals. The optimal order of interference cancellation is detecting the user with the most power allocation to the user with the least power allocation \cite{Ref_jiang20236GCH10}. We write $\rho_k=(\mathbf{g}_k^T \boldsymbol{\Theta}^{\star}_{\hat{k}} \mathbf{H} +\mathbf{f}_k^T)\mathbf{w}_k^\star$, $\forall k$ to denote the effective gain of the combined channel for user $k$. Without loss of generality, assume that user $1$ has the largest combined channel gain, and user $K$ is the weakest, i.e., 
$\| \rho_1 \|^2 \geqslant \|\rho_2\|^2 \geqslant \ldots \geqslant \|\rho_K\|^2$.
 With this order, each NOMA-IRS user first decodes $s_K$, and then subtracts its resultant component from the received signal. As a result, a typical user $k$ after the first SIC iteration gets 
\begin{equation}
    \tilde{y}_k = y_k - \rho_k \sqrt{\alpha_K P_d}  s_K = \rho_k\sum_{k=1}^{K-1} \sqrt{\alpha_k P_d} s_k+n_k.
\end{equation}
In the second iteration, the user decodes $s_{K-1}$ using the remaining signal $ \tilde{y}_k$. The cancellation iterates until each user gets the symbol intended for it.  User $k$ successfully cancels the signals from user $k+1$ to $K$ but suffers from the interference from user $1$ to $k-1$. Treating uncancelled multi-user interference as noise, the received SNR for user $k$ is
\begin{equation}
    \gamma_k=\frac{\|\rho_k\|^2 \alpha_k P_d}{ \|\rho_k\|^2 \sum_{k'=1}^{k-1}\alpha_{k'} P_d + \sigma_n^2 }.
\end{equation}
Thus, the sum rate of NOMA-based IRS is figured out as
\begin{equation}
    C_{noma}=\sum_{k=1}^K \log \left ( 1+\frac{\|\rho_k\|^2\alpha_k P_d}{\|\rho_k\|^2 \sum_{k'=1}^{k-1}\alpha_{k'} P_d + \sigma_n^2 }   \right).
\end{equation}

\subsection{SDMA}  
When the elements of an antenna array are highly correlated, it enables traditional angle-based beamforming. Directional beams are steered toward specific users, allowing multiple users to share the same frequency band simultaneously. In a macro-cell scenario, scatterers are positioned near users, and the scatterer size is significantly smaller than the BS-UE distance. Consequently, the angle of departure (AOD) for various signal paths from a user in the far field tends to be approximately the same, denoted by $\theta$. For a uniform linear array (ULA) with inter-antenna spacing set at half a wavelength, the steering vector can be expressed as \cite{Ref_yang2011random}
\begin{equation}
    \textbf{a}(\theta) = \left[1,e^{-j\pi \sin(\theta)},\ldots,e^{-j(M-1)\pi \sin(\theta)}   \right]^T.
\end{equation} 
Let the first antenna $m=1$ of the array be the reference point, \eqref{EQN_f_k} can be rewritten as $\mathbf{f}_{k}=f_{k1} \textbf{a}(\theta)$, $\forall k=1,\ldots,K$, while $\mathbf{h}_{n}=h_{n1} \textbf{a}(\theta)$, $\forall n=1,\ldots,N$.

The rationale of SDMA-based IRS is that the BS generates independent beams toward the surface and users to deal with inter-user interference. The beam pointing to the surface can be referred to as a reflected beam. Without loss of generality, a specific user $\hat{k}$ can be chosen for this reflected beam, while other users utilize a direct beam to communicate with the BS \cite{Ref_jiang2024beam}. 
As the positions of the BS and IRS remain fixed, the angle of IRS denoted by $\theta_I$ is readily determinable and remains constant over an extended period. The optimal weighting vector to generate a beam towards the IRS is $\mathbf{w}_I=\sqrt{1/M}\textbf{a}^H(\theta_I)$, resulting in the maximal gain
\begin{equation} \label{optimalWeight1}
    B(\theta_I)=\sqrt{\frac{1}{M}} \textbf{a}^H(\theta_I)  \textbf{a}(\theta_I)=\sqrt{M}.
\end{equation} 

For a typical non-IRS-aided user with the AOD of $\theta_{k}$,  the optimal weighting vector is $\mathbf{w}_{k}=\sqrt{\frac{1}{M}}\textbf{a}^H(\theta_{k})$. Thus, the direct beam gain is  
\begin{equation}
    B_{k}(\theta_{k})=\sqrt{\frac{1}{M}} \textbf{a}^H(\theta_{k})  \textbf{a}(\theta_{k})=\sqrt{M}.
\end{equation}
The transmitted vector is given by $\mathbf{x}=\sum_{k=1}^K \mathbf{w}_{k} s_k$. Substituting it into \eqref{EQN_IRS_RxSignal_Matrix}, we obtain the received signal of user $\hat{k}$ as
\begin{align} \label{XXXX} \nonumber
    y_{\hat{k}} &= \sqrt{MP_d}\mathbf{g}_{\hat{k}}^T \boldsymbol{\Theta} \bar{\mathbf{h}}_1 s_{\hat{k}} +n_{\hat{k}}\\
     &= \sqrt{MP_d}\left( \sum_{n=1}^N g_{n{\hat{k}}} e^{j\phi_n} h_{n1} \right)  s_{\hat{k}} +n_{\hat{k}},
\end{align}
where $\bar{\mathbf{h}}_1\in \mathbb{C}^{N\times 1}$ is the first column of $\mathbf{H}$. Due to the spatial isolation, the reflection optimization of user $\hat{k}$ is equivalent to that of a single-user system, simplifying the joint optimization of active/passive beamforming
Therefore, the optimal phase shift for reflecting element $n$ equals to 
\begin{equation} \label{IRSeqn:optimalphase}    
\phi_n^\star = \mod \Bigl[\psi_a-\arg(h_{n1})-\arg(g_{n\hat{k}}), 2\pi   \Bigr],
\end{equation}
where $\psi_a$ stands for an arbitrary phase value. It implies that the phase shift of each reflected signal is compensated, such that the residual phase of each branch is equal to $\psi_a$, for coherent combining at the receiver.

Thus, \eqref{XXXX} equals to
\begin{equation} 
    y_{\hat{k}}= \sqrt{MP_d}  \sum_{n=1}^N |g_{n{\hat{k}}} h_{n1}|e^{j\psi_a}   s_{\hat{k}} +n_{\hat{k}}.
\end{equation}
Similarly, we know the received signal for user $k\neq \hat{k}$ is
\begin{equation} \label{}
    y_k= \sqrt{MP_d}f_{k1} s_{k} +n_k.
\end{equation}
Then, the sum rate of the SDMA-based IRS system is
 \begin{align} \nonumber
     C_{sdma}&= \log\left(1+\frac{M P_d |\sum_{n=1}^N |g_{n{\hat{k}}} h_{n1}||^2}{\sigma_n^2} \right)  \\ \nonumber
     &+\sum_{k\neq \hat{k}}   \log\left(1+\frac{M P_d |f_{k1}|^2}{\sigma_n^2} \right).
  \end{align}

\subsection{Opportunistic User Selection}
Because of the shared transmission resource, a trade-off exists in a multi-user system: if someone wishes for a higher rate, it necessitates that other users reduce their rates. Hence, the appropriate performance metric is the sum capacity 
\begin{equation} \label{EQN_SumRate}
    C =\sum_{k=1}^K \log\left(1+\frac{ \Bigl| \left(\mathbf{g}_{k}^T \boldsymbol{\Theta} \mathbf{H} +\mathbf{f}_k^T\right)\mathbf{w}_k  \Bigr|^2 P_d}{\sigma_n^2} \right).
\end{equation}
It raises the optimization formula:
\begin{equation}  
\begin{aligned} \label{eqnIRS:optimizationMRTvector}
\max_{\boldsymbol{\Theta},\:\mathbf{w}_k}\quad &  \sum_{k=1}^K \log\left(1+\frac{ \Bigl| \left(\mathbf{g}_{k}^T \boldsymbol{\Theta} \mathbf{H} +\mathbf{f}_k^T\right)\mathbf{w}_k  \Bigr|^2 P_d}{\sigma_n^2} \right) \\
\textrm{s.t.} \quad & \|\mathbf{w}_k\|^2= 1,\: \forall k\\ 
  \quad & \phi_n\in [0,2\pi), \: \forall n=1,\ldots,N, 
\end{aligned}
\end{equation}
which is non-convex due to the lack of joint concavity in the objective function concerning both $\boldsymbol{\Theta}$ and $\mathbf{w}_k$. Hence, the optimization of passive beamforming is unfeasible. Consequently, there is a need to explore an innovative approach to simplify the system design.  

In wireless systems, opportunistic communications \cite{Ref_jiang2016robust} strike a favorable blance between performance and complexity by leveraging multi-user diversity gain. The selection of an opportunistic user, identified as $k^\star$, is based on the best channel condition \cite{Ref_jiang2023userSelection} according to the expression: 
\begin{equation}
    k^\star=\arg \max_{k\in\{1,\ldots,K\}} \left\{\Bigl| \left(\mathbf{g}_{k}^T \boldsymbol{\Theta} \mathbf{H} +\mathbf{f}_k^T\right)\mathbf{w}_k  \Bigr|^2 \right\}.
\end{equation}
In this scenario, the BS transmits the symbol exclusively to $k^\star$, effectively degrading the multi-user system to a single-user system. Consequently, the sum capacity in \eqref{EQN_SumRate} becomes
\begin{equation} \label{EQN_IRS_MTRcapacity}
  C = \log  \left(  1+\Bigl\| \mathbf{g}_{k^\star}^T \boldsymbol{\Theta} \mathbf{H} +\mathbf{f}_{k^\star}^T  \Bigr\|^2 \frac{P_d}{\sigma_n^2}  \right).
\end{equation} 
The application of SDR has been employed to address a non-convex quadratically constrained quadratic program (QCQP) in a single-user IRS system, as discussed in \cite{Ref_wu2019intelligent}. However, its direct applicability to the multi-user IRS system is not possible. Through user selection, SDR can be employed to jointly optimize the active and passive beamforming specifically for the selected user.

Define $\mathbf{q}=\left[q_{1},q_{2},\ldots,q_N\right]^H$, where $q_{n}=e^{j\phi_{n}}$. Let  $\boldsymbol{\chi}=\mathrm{diag}(\mathbf{g}_{k^\star}^T) \mathbf{H} \in \mathbb{C}^{N\times M}$, we have $\mathbf{g}_{k^\star}^T \boldsymbol{\Theta} \mathbf{H} =\mathbf{q}^H\boldsymbol{\chi}\in \mathbb{C}^{1\times M} $.
Thus, $\left\|\mathbf{g}_{k^\star}^T \boldsymbol{\Theta} \mathbf{H}  +\mathbf{f}_{k^\star}^T\right\|^2=\left\|\mathbf{q}^H\boldsymbol{\chi} +\mathbf{f}_{k^\star}^T\right\|^2$.   
 Introducing an auxiliary variable $t$, we have the optimization problem:
\begin{equation} \begin{aligned} \label{eqn_IRS_relaxedOptimization}
    \max_{\mathbf{q}}  \quad & \left\|t\mathbf{q}^H\boldsymbol{\chi} +\mathbf{f}_{k^\star}^T\right\|^2\\
     =\max_{\mathbf{q}} \quad & t^2\mathbf{q}^H\boldsymbol{\chi}\boldsymbol{\chi}^H\mathbf{q}+t\mathbf{q}^H\boldsymbol{\chi}\mathbf{f}_{k^\star}^*+t\mathbf{f}_{k^\star}^T\boldsymbol{\chi}^H\mathbf{q}+\|\mathbf{f}_{k^\star}\|^2.
     \end{aligned}
\end{equation}
Define $\mathbf{C}=\begin{bmatrix}\boldsymbol{\chi}\boldsymbol{\chi}^H&\boldsymbol{\chi}\mathbf{f}_{k^\star}^H\\ \mathbf{f}_{k^\star}\boldsymbol{\chi}^H& \|\mathbf{f}_{k^\star}\|^2\end{bmatrix},\:\:\mathbf{v}= \begin{bmatrix}\mathbf{q}\\ t\end{bmatrix}$, and $\mathbf{V}=\mathbf{v}\mathbf{v}^H$, we have $\mathbf{v}^H\mathbf{C}\mathbf{v}=\mathrm{Tr}(\mathbf{C}\mathbf{V})$, where $\mathrm{Tr}(\cdot)$ denotes the trace of a matrix.
Thus, \eqref{eqn_IRS_relaxedOptimization} is reformulated as 
\begin{equation}  \label{RIS_EQN_optimizationTrace}
\begin{aligned} \max_{\mathbf{V}}\quad &  \mathrm{Tr} \left(  \mathbf{C}\mathbf{V} \right)\\
\textrm{s.t.}  \quad & \mathbf{V}_{m,m}=1, \: \forall m=1,\ldots,M \\
  \quad & \mathbf{V}\succ 1
\end{aligned},
\end{equation}
where $\mathbf{V}_{m,m}$ means the $m^{th}$ diagonal element of $\mathbf{V}$, and $\succ$ stands for a positive semi-definite matrix.
The optimization finally becomes a semi-definite program, whose globally optimal solution $\mathbf{V}^\star$ can be got by a numerical algorithm named CVX \cite{cvx}.

A sub-optimal solution for \eqref{RIS_EQN_optimizationTrace} is given by $\bar{\mathbf{v}}=\mathbf{U}\boldsymbol{\Sigma}^{1/2}\mathbf{r}$,
where $\mathbf{r}\in \mathcal{CN}(\mathbf{0},\mathbf{I}_{N+1})$ is a Gaussian RV, a unitary matrix $\mathbf{U}$ and a diagonal matrix $\boldsymbol{\Sigma}$ are obtained from the eigenvalue decomposition $\mathbf{V}^\star=\mathbf{U}\boldsymbol{\Sigma}\mathbf{U}^H$. The jointly optimized phase shifts are determined as $\boldsymbol{\Theta}^\circledast =\mathrm{diag} \left\{e^{j \arg\left( \left[\frac{\bar{\mathbf{v}}}{\bar{v}_{_{N+1}}} \right]_{1:N}\right)}\right\}$, where $[\cdot]_{1:N}$ denotes a sub-vector extracting the first $N$ elements and $\bar{v}_{_{N+1}}$ is the last element of $\bar{\mathbf{v}}$. Applying $k=k^\star$ and $\boldsymbol{\Theta}^\circledast$ into \eqref{EQN_IRS_MTRcapacity} yields a sum capacity 
\begin{equation}
  C_{us} = \log  \left(  1+\Bigl\| \mathbf{g}_{k^\star}^T \boldsymbol{\Theta}^\circledast \mathbf{H} +\mathbf{f}_{k^\star}^T  \Bigr\|^2 \frac{P_d}{\sigma_n^2}  \right).
\end{equation}

\subsection{Opportunistic Beamforming}

Many existing approaches optimizing IRS phase shifts assume perfect CSI, which is challenging to acquire due to limited signal processing capability at the IRS. Recent works on channel estimation protocols for such systems reveal that training time increases with the number of reflecting elements, compromising performance gains from deploying a large number of IRS elements due to high feedback overhead. Opportunistic beamforming \cite{Ref_nadeem2021opportunistic} in a random-rotation-based IRS-aided multi-user system involves beam selection during a training period with multiple slots. In this period, the BS generates multiple sets of orthonormal beamforming vectors. Each user provides feedback only on its highest SNR. The BS then employs beam selection based on the highest sum rate.

At the beginning of each time frame, there is a training period comprising $T$ training slots. Within this training period, the BS generates multiple sets of $B$ orthonormal beamforming vectors, with $B<K$. At each training slot, denoted as $t=1,\ldots,T$, $B$ pilot symbols are multiplexed by their corresponding beamforming vectors. Specifically, the transmitted vector  during the $t^{th}$ training slot is expressed by $\mathbf{x}_t=\sum_{b=1}^B \mathbf{w}_{b,t} s_b$, where $s_b$ is the $b^{th}$ pilot symbol \cite{Ref_dimitropoulou2022opportunistic}. 
Substituting $\mathbf{x}_t$ into \eqref{EQN_IRS_RxSignal_Matrix}, we obtain
\begin{equation} 
    y_k= \sqrt{P_d}\Bigl(\mathbf{g}_k^T \boldsymbol{\Theta}_r \mathbf{H} +\mathbf{f}_k^T\Bigr)\sum_{b=1}^B \mathbf{w}_{b,t} s_b +n_k,
\end{equation}
recalling $\boldsymbol{\Theta}_{r}$, each entry of which takes value randomly from $[0,2\pi)$.
During each training slot $t$, every user $k$ calculates the values of SNRs per beam by considering $s_b$, where $b$ takes values from $1$ to $B$, as the desired pilot symbol, while treating the remaining symbols as undesirable noise. Thus, we have
\begin{equation}
    \gamma_{k}[t,b]=\frac{ P_d \left \|\left(\mathbf{g}_k^T \boldsymbol{\Theta}_r \mathbf{H} +\mathbf{f}_k^T\right)\mathbf{w}_{b,t} \right\|^2  }{ P_d \sum_{b'\neq b} \left \|\left(\mathbf{g}_k^T \boldsymbol{\Theta}_r \mathbf{H} +\mathbf{f}_k^T\right)\mathbf{w}_{b',t} \right\|^2 + \sigma_n^2 }.
\end{equation}

Then, each user provides feedback on its maximal SNR represented as $\gamma^\star[t,k]=\max_{b=1,\ldots,B}(\gamma_{k}[t,b])$, along with the corresponding index $b_k^\star=\arg \max_{b=1,\ldots,B}(\gamma_{k}[t,b])$. The BS identifies the best user for each beam,  and therefore a set of $B$ users. We write $\gamma^\star[t,b]$ to denote the SNR of the best user for beam $b$ at training slot $t$. Using the scheduled users got at training slot $t$, an achievable sum rate equals 
\begin{equation}
    R_t= \sum_{b=1}^B \log \left( 1+ \gamma^\star[t,b]   \right).
\end{equation}
Following the training period, there are $T$ sets, each comprising $B$ users. The BS then chooses the set of $B$ users that yields the highest sum-rate $\max_{t\in\{1,\ldots,T\}} (R_t)$. Consequently, during the subsequent data transmission period, the data symbols are transmitted through the IRS to the respective users and the achievable sum rate is $ C= \max_{t\in\{1,\ldots,T\}} \sum_{b=1}^B \log \left( 1+ \gamma^\star[t,b]   \right)$.

\section{Conclusions}
This paper reviewed various multiple access techniques in IRS-aided wireless systems. We comparatively presented the fundamentals of different MA approaches in a unified system model. The functionalities of these techniques were illustrated through mathematical representations of transmitted and received signals, alongside the derivation of closed-form expressions for achievable spectral efficiency and sum rate. The primary motivation of our work resides in delivering an exhaustive survey, presenting a unified illustration, and providing a holistic view. Our goal is to facilitate researchers a comprehensive understanding of challenges and existing solutions, and offer valuable insights to foster the design of efficient multiple access strategies for IRS-aided systems.

\bibliographystyle{IEEEtran}
\bibliography{IEEEabrv,Ref_eucnc}

\begin{thebibliography}{10}
\providecommand{\url}[1]{#1}
\csname url@samestyle\endcsname
\providecommand{\newblock}{\relax}
\providecommand{\bibinfo}[2]{#2}
\providecommand{\BIBentrySTDinterwordspacing}{\spaceskip=0pt\relax}
\providecommand{\BIBentryALTinterwordstretchfactor}{4}
\providecommand{\BIBentryALTinterwordspacing}{\spaceskip=\fontdimen2\font plus
\BIBentryALTinterwordstretchfactor\fontdimen3\font minus \fontdimen4\font\relax}
\providecommand{\BIBforeignlanguage}[2]{{%
\expandafter\ifx\csname l@#1\endcsname\relax
\typeout{** WARNING: IEEEtran.bst: No hyphenation pattern has been}%
\typeout{** loaded for the language `#1'. Using the pattern for}%
\typeout{** the default language instead.}%
\else
\language=\csname l@#1\endcsname
\fi
#2}}
\providecommand{\BIBdecl}{\relax}
\BIBdecl

\bibitem{Ref_renzo2020smart}
M.~D. Renzo \emph{et~al.}, ``Smart radio environments empowered by reconfigurable intelligent surfaces: How it works, state of research, and the road ahead,'' \emph{{IEEE} J. Sel. Areas Commun.}, vol.~38, no.~11, pp. 2450 -- 2525, Nov. 2020.

\bibitem{Ref_jiang2024TextBook}
W.~Jiang and B.~Han, \emph{Cellular Communication Networks and Standards: The Evolution from {1G} to {6G}}.\hskip 1em plus 0.5em minus 0.4em\relax Cham, Switzerland: Springer, 2024.

\bibitem{Ref_jiang20236GCH7}
W.~Jiang and F.-L. Luo, ``Intelligent reflecting surface-aided communications for {6G},'' in \emph{{6G} Key Technologies: A Comprehensive Guide}.\hskip 1em plus 0.5em minus 0.4em\relax New York, USA: John Wiley\&Sons and IEEE Press, 2023, ch.~7.

\bibitem{Ref_zhang2023activeRIS}
Z.~Zhang \emph{et~al.}, ``Active {RIS} vs. passive {RIS}: Which will prevail in {6G}?'' \emph{{IEEE} Trans. Commun.}, vol.~71, no.~3, pp. 1707 -- 1725, Mar. 2023.

\bibitem{Ref_jiang2021road}
W.~Jiang \emph{et~al.}, ``The road towards {6G}: A comprehensive survey,'' \emph{IEEE Open J. Commun. Society}, vol.~2, pp. 334--366, Feb. 2021.

\bibitem{Ref_wu2019intelligent}
Q.~Wu and R.~Zhang, ``Intelligent reflecting surface enhanced wireless network via joint active and passive beamforming,'' \emph{{IEEE} Trans. Wireless Commun.}, vol.~18, no.~11, pp. 5394 -- 5409, Nov. 2019.

\bibitem{Ref_wang2020channel}
Z.~Wang, L.~Liu, and S.~Cui, ``Channel estimation for intelligent reflecting surface assisted multiuser communications: Framework, algorithms, and analysis,'' \emph{{IEEE} Trans. Wireless Commun.}, vol.~19, no.~10, pp. 6607 -- 6620, Oct. 2020.

\bibitem{Ref_jiang2023performance}
W.~Jiang and H.~Schotten, ``Performance impact of channel aging and phase noise on intelligent reflecting surface,'' \emph{{IEEE} Commun. Lett.}, vol.~27, no.~1, pp. 347--351, Jan. 2023.

\bibitem{Ref_jiang20236GCH8}
W.~Jiang and F.-L. Luo, ``Multiple dimensional and antenna techniques for {6G},'' in \emph{{6G} Key Technologies: A Comprehensive Guide}.\hskip 1em plus 0.5em minus 0.4em\relax New York, USA: John Wiley\&Sons and IEEE Press, 2023, ch.~8.

\bibitem{Ref_zheng2020intelligent_COML}
B.~Zheng, Q.~Wu, and R.~Zhang, ``Intelligent reflecting surface-assisted multiple access with user pairing: {NOMA or OMA}?'' \emph{{IEEE} Commun. Lett.}, vol.~24, no.~4, pp. 753 -- 757, Apr. 2020.

\bibitem{Ref_jiang2023userscheduling}
W.~Jiang and H.~Schotten, ``User scheduling and passive beamforming for {FDMA/OFDMA} in intelligent reflection surface,'' in \emph{Proc. 2023 IEEE 97th Veh. Techno. Conf. (VTC2023-Spring)}, Florence, Italy, Jun. 2023.

\bibitem{Ref_chu2022resource}
Z.~Chu \emph{et~al.}, ``Resource allocation for {IRS}-assisted wireless-powered {FDMA IoT} networks,'' \emph{IEEE Internet of Things J.}, vol.~9, no.~11, pp. 8774 -- 8785, Jun. 2022.

\bibitem{Ref_jiang2023orthogonal}
W.~Jiang and H.~Schotten, ``Orthogonal and non-orthogonal multiple access for intelligent reflection surface in {6G} systems,'' in \emph{Proc. 2023 IEEE Wireless Commun. and Netw. Conf. (WCNC)}, Glasgow, Scotland, UK, Mar. 2023.

\bibitem{Ref_ding2020simple}
Z.~Ding and H.~V. Poor, ``A simple design of {IRS-NOMA} transmission,'' \emph{{IEEE} Commun. Lett.}, vol.~24, no.~5, pp. 1119 -- 1123, May 2020.

\bibitem{Ref_jiang2024beam}
W.~Jiang and H.~D. Schotten, ``Beam-based multiple access for {IRS}-aided millimeter-wave and terahertz communications,'' in \emph{Proc. 2024 IEEE Wireless Commun. and Netw. Conf. (WCNC)}, Dubai, UAE, Apr. 2024.

\bibitem{Ref_jiang2022dualbeam}
W.~Jiang and H.~Schotten, ``Dual-beam intelligent reflecting surface for millimeter and {THz} communications,'' in \emph{Proc. 2022 IEEE 95th Veh. Techno. Conf. (VTC2022-Spring)}, Helsinki, Finland, Jun. 2022.

\bibitem{Ref_jiang2023capacity}
W.~Jiang and H.~D. Schotten, ``Capacity analysis and rate maximization design in {RIS}-aided uplink multi-user {MIMO},'' in \emph{Proc. 2023 IEEE Wireless Commun. and Netw. Conf. (WCNC)}, Glasgow, Scotland, UK, Mar. 2023.

\bibitem{Ref_gan2021user}
X.~Gan \emph{et~al.}, ``User selection in reconfigurable intelligent surface assisted communication systems,'' \emph{{IEEE} Commun. Lett.}, vol.~25, no.~4, pp. 1353 -- 1357, Apr. 2021.

\bibitem{Ref_jiang2023userSelection}
W.~Jiang and H.~Schotten, ``User selection for simple passive beamforming in multi-{RIS}-aided multi-user communications,'' in \emph{Proc. 2023 IEEE 97th Veh. Techno. Conf. (VTC2023-Spring)}, Florence, Italy, Jun. 2023.

\bibitem{Ref_nadeem2021opportunistic}
Q.-U.-A. Nadeem \emph{et~al.}, ``Opportunistic beamforming using an intelligent reflecting surface without instantaneous {CSI},'' \emph{IEEE Wireless Commun. Lett.}, vol.~10, no.~1, pp. 146 -- 150, Jan. 2021.

\bibitem{Ref_dimitropoulou2022opportunistic}
M.~Dimitropoulou, C.~Psomas, and I.~Krikidis, ``Opportunistic beamforming with beam selection in {IRS}-aided communications,'' in \emph{Proc. 2022 IEEE Int. Commun. Conf. (ICC)}, Seoul, South Korea, May 2022.

\bibitem{Ref_jiang2023opportunistic}
W.~Jiang and H.~Schotten, ``Opportunistic reflection in reconfigurable intelligent surface-assisted wireless networks,'' in \emph{Proc. 2023 IEEE Int. Symp. on Pers., Indoor and Mobile Radio Commun. (PIMRC)}, Toronto, Canada, Sep. 2023.

\bibitem{Ref_jiang2023simple}
W.~Jiang and H.~D. Schotten, ``A simple multiple-access design for reconfigurable intelligent surface-aided systems,'' in \emph{Proc. 2023 IEEE Global Commun. Conf. (Globecom)}, Kuala Lumpur, Malaysia, Dec. 2023.

\bibitem{Ref_jiang2022multiuser}
W.~Jiang and H.~Schotten, ``Multi-user reconfigurable intelligent surface-aided communications under discrete phase shifts,'' in \emph{Proc. 36th IEEE Int. Workshop on Commun. Qual. and Reliability (CQR 2022)}, Arlington, United States, Sep. 2022.

\bibitem{Ref_jiang2022intelligent}
W.~Jiang and H.~D. Schotten, ``Intelligent reflecting vehicle surface: A novel {IRS} paradigm for moving vehicular networks,'' in \emph{Proc. 2022 IEEE 40th Military Commun. Conf. (MILCOM 2022)}, Rockville, MA, USA, Nov. 2022.

\bibitem{Ref_wu2020beamforming}
Q.~Wu and R.~Zhang, ``Beamforming optimization for wireless network aided by intelligent reflecting surface with discrete phase shifts,'' \emph{{IEEE} Trans. Commun.}, vol.~68, no.~3, pp. 838 -- 1851, Mar. 2020.

\bibitem{Ref_zhang2018space}
L.~Zhang \emph{et~al.}, ``Space-time-coding digital metasurfaces,'' \emph{Nature Commun.}, vol.~9, no.~1, p. 4334, 1998.

\bibitem{Ref_jiang20236GCH10}
W.~Jiang and F.-L. Luo, ``Adaptive and non-orthogoanal multiple access systems in {6G},'' in \emph{{6G} Key Technologies: A Comprehensive Guide}.\hskip 1em plus 0.5em minus 0.4em\relax New York, USA: John Wiley\&Sons and IEEE Press, 2023, ch.~10.

\bibitem{Ref_yang2011random}
X.~Yang, W.~Jiang, and B.~Vucetic, ``A random beamforming technique for broadcast channels in multiple antenna systems,'' in \emph{Proc. 2011 IEEE Veh. Techno. Conf. (VTC Fall)}, San Francisco, USA, Sep. 2011.

\bibitem{Ref_jiang2016robust}
W.~Jiang, T.~Kaiser, and A.~J.~H. Vinck, ``A robust opportunistic relaying strategy for co-operative wireless communications,'' \emph{{IEEE} Trans. Wireless Commun.}, vol.~15, no.~4, pp. 2642--2655, Apr. 2016.

\bibitem{cvx}
M.~Grant and S.~Boyd, ``{CVX}: Matlab software for disciplined convex programming, version 2.1,'' \url{http://cvxr.com/cvx}, Mar. 2014.

\end{thebibliography}

\end{document}